\documentstyle[12pt]{article}
\newlength{\extraspace}
\newlength{\extraspaces}
\newcommand{\be}{\begin{equation}
\addtolength{\abovedisplayskip}{\extraspaces}
\addtolength{\belowdisplayskip}{\extraspaces}
\addtolength{\abovedisplayshortskip}{\extraspace}
\addtolength{\belowdisplayshortskip}{\extraspace}}
\newcommand{\ee}{\end{equation}}

\newcommand{\ba}{\begin{eqnarray}
\addtolength{\abovedisplayskip}{\extraspaces}
\addtolength{\belowdisplayskip}{\extraspaces}
\addtolength{\abovedisplayshortskip}{\extraspace}
\addtolength{\belowdisplayshortskip}{\extraspace}}
\newcommand{\ea}{\end{eqnarray}}

\newcommand{\newsection}[1]{
\vspace{15mm}
\pagebreak[3]
\addtocounter{section}{1}
\setcounter{equation}{0}
\setcounter{subsection}{0}
\setcounter{footnote}{0}
\begin{flushleft}
{\large\bf \thesection. #1}
\end{flushleft}
\nopagebreak
\medskip
\nopagebreak}

\newcommand{\Tr}{{\rm Tr}}
\newcommand{\Dmrns}{{\cal D}_{\mu\rho,\nu\sigma}}

\begin{document}

\addtolength{\baselineskip}{.8mm}

{\thispagestyle{empty}
\noindent \hspace{1cm}  \hfill March 1996 \hspace{1cm}\\
\mbox{}                 \hfill IFUP--TH 14/96 \hspace{1cm}\\
\mbox{}                 \hfill UCY--PHY--96/6 \hspace{1cm}\\

\begin{center}\vspace*{1.0cm}
{\large\bf Gauge--invariant field correlators in QCD}\\
{\large\bf at finite temperature
 \footnote{Partially supported by MURST (Italian Ministry of the University 
 and of Scientific and Technological Research) and by the EC contract 
 CHEX--CT92--0051.} }\\
\vspace*{1.0cm}
{\large A. Di Giacomo, E. Meggiolaro}\\
\vspace*{0.5cm}{\normalsize
{Dipartimento di Fisica, \\
Universit\`a di Pisa, \\ 
and INFN, Sezione di Pisa,\\
I--56100 Pisa, Italy.}}\\
\vspace*{1.0cm}
{\large H. Panagopoulos}\\
\vspace*{0.5cm}{\normalsize
{Department of Natural Sciences,\\
University of Cyprus,\\
1678 Nicosia, Cyprus.}}\\
\vspace*{2cm}{\large \bf Abstract}
\end{center}
\noindent
We study by numerical simulations on a lattice the behaviour of 
the gauge--invariant two--point correlation functions of the gauge field 
strengths across the deconfinement phase transition.
}
\vfill\eject

\newsection{Introduction}

\noindent
Gauge--invariant correlation functions of the field strenghts in the QCD 
vacuum play an important role in high--energy phenomenology and in models 
of stochastic confinement \cite{Dosch87,Dosch88,Simonov89}.
Recently, a determination of such correlators at zero temperature has been 
done on the lattice, in a range of distances from $0.1$ to 1 fm 
\cite{DiGiacomo96}.
The technique used to make the computation feasible is a local cooling of 
the configurations: this procedure freezes local fluctuations, leaving 
long--range correlations unchanged.
In this paper we determine the behaviour of the correlators at finite 
temperature for the pure--gauge theory with $SU(3)$ colour group and in 
particular we study their behaviour across the deconfining phase transition.
The motivations to do that stem from Refs. 
\cite{Simonov1,Simonov2,Simonov3}.

\newsection{Notation}

\noindent
To simulate the system at finite temperature, a lattice is used of spatial 
extent $N_\sigma \gg N_\tau$, $N_\tau$ being the temporal extent, 
with periodic boundary conditions.
The temperature $T$ corresponding to a given value of $\beta = 6/g^2$ is 
given by
\be
N_\tau \cdot a(\beta) = {1 \over T} ~,
\ee
where $a(\beta)$ is the lattice spacing.
From renormalization group arguments,
\be
a(\beta) = {1\over\Lambda_L} f(\beta) ~,
\ee
where $\Lambda_L$ is the fundamental constant of QCD in the lattice 
renormalization scheme: its value, extracted from the string tension 
\cite{Michael88}, turns out to be about 5 MeV.
At large enough $\beta$, $f(\beta)$ is given by the usual two--loop 
expression:
\be
f(\beta) = \left({8\over33}\,\pi^2\beta\right)
^{ 51/121 } \exp\left(-{4\over33}\pi^2\beta\right)
\left[1+{\cal O}(1/\beta)\right] ~,
\ee
for gauge group $SU(3)$ and in the absence of quarks.

The gauge--invariant two--point correlators of the field strengths in the 
QCD vacuum are defined as
\be
\Dmrns(x) = \langle 0| 
\Tr \left\{ G_{\mu\rho}(x) S(x,0) G_{\nu\sigma}(0) S^\dagger(x,0) \right\}
|0\rangle ~,
\ee
where $G_{\mu\rho} = gT^aG^a_{\mu\rho}$ is the field--strength tensor 
and $T^a$ are the generators of the colour gauge group in the fundamental 
representation. Moreover,
\be
S(x,0) = {\rm P}\exp\left(i\int^1_0dt\,x^\mu A_\mu(xt)\right) ~,
\ee
with $A_\mu=gT^aA^a_\mu$, is the Schwinger phase operator needed to 
parallel--transport the tensor $G_{\nu\sigma}(0)$ to the point $x$, 
so to make $\Dmrns(x)$ a gauge--invariant quantity.
At zero temperature, that is on a symmetric lattice $N_\sigma = N_\tau$,
they are expressed in terms of two independent invariant functions of 
$x^2$, called ${\cal D} (x^2)$ and ${\cal D}_1 (x^2)$, as follows
\cite{Dosch87,Dosch88,Simonov89}:
\ba
\lefteqn{
\Dmrns(x) = (g_{\mu\nu}g_{\rho\sigma} - g_{\mu\sigma}g_{\rho\nu})
\left[ {\cal D}(x^2) + {\cal D}_1(x^2) \right] } \nonumber \\
& & + (x_\mu x_\nu g_{\rho\sigma} - x_\mu x_\sigma g_{\rho\nu} 
+ x_\rho x_\sigma g_{\mu\nu} - x_\rho x_\nu g_{\mu\sigma})
{\partial{\cal D}_1(x^2) \over \partial x^2} ~.
\ea
At finite temperature, the $O(4)$ space--time symmetry is broken down to
the spatial $O(3)$ symmetry and in principle the bilocal correlators are now
expressed in terms of five independent functions
\cite{Simonov1,Simonov2,Simonov3}; two of them are needed to
describe the electric--electric correlations:
\ba
\lefteqn{
\langle 0| \Tr \left\{ E_i (x) S(x,y) E_k (y) S^\dagger(x,y) \right\}
|0\rangle } \nonumber \\
& & = \delta_{ik} \left[ D^E + D_1^E + u_4^2 {\partial D_1^E \over 
\partial u_4^2} \right] + u_i u_k {\partial D_1^E \over 
\partial \vec{u}^2} ~,
\ea
where $E_i = G_{i4}$ is the electric field operator and 
$u_\mu = x_\mu - y_\mu$ [$\vec{u}^2 = (\vec{x} - \vec{y})^2$].
We are considering the Euclidean theory.

Two further functions are needed for the magnetic--magnetic correlations:
\ba
\lefteqn{
\langle 0| \Tr \left\{ B_i (x) S(x,y) B_k (y) S^\dagger(x,y) \right\}
|0\rangle } \nonumber \\
& & = \delta_{ik} \left[ D^B + D_1^B + \vec{u}^2 {\partial D_1^B \over 
\partial \vec{u}^2} \right] - u_i u_k {\partial D_1^B \over 
\partial \vec{u}^2} ~,
\ea
where $B_k = {1 \over 2} \varepsilon_{ijk} G_{ij}$ is the magnetic field
operator.

Finally, one more function is necessary to describe the mixed 
electric--magnetic correlations:
\be
\langle 0| \Tr \left\{ E_i (x) S(x,y) B_k (y) S^\dagger(x,y) \right\}
|0\rangle = -{1 \over 2} \varepsilon_{ikn} u_n
{\partial D_1^{BE} \over \partial u_4} ~.
\ee
In Eqs. (2.7), (2.8) and (2.9), the five quantities $D^E$, $D_1^E$, $D^B$, 
$D_1^B$ and $D_1^{BE}$ are all functions of $\vec{u}^2$, due to rotational 
invariance, and of $u_4^2$, due to time--reversal invariance.

From the conclusions of Refs. \cite{Simonov1,Simonov2,Simonov3}, 
one expects that $D^E$ is related to the 
string tension and should have a drop just above the deconfinement critical 
temperature $T_c$. In other words, $D^E$ is expected to be the order 
parameter of the confinement; on the contrary,
$D^E_1$ does not contribute to the area law of the temporal Wilson loop.
(Similarly, $D^B_1$ does not contribute to the area law of the spatial 
Wilson loop.)

\newsection{Results}

\noindent
We have determined the following four quantities
\ba
{\cal D}_\parallel^E (\vec{u}^2,0) &\equiv&
 {\cal D}^E (\vec{u}^2,0) + {\cal D}_1^E (\vec{u}^2,0) + 
\vec{u}^2 {\partial{\cal D}_1^E \over \partial \vec{u}^2} (\vec{u}^2,0) ~; 
\nonumber \\
{\cal D}_\perp^E (\vec{u}^2,0) &\equiv& {\cal D}^E (\vec{u}^2,0) 
+ {\cal D}_1^E (\vec{u}^2,0) ~; 
\nonumber \\
{\cal D}_\parallel^B (\vec{u}^2,0) &\equiv&
 {\cal D}^B (\vec{u}^2,0) + {\cal D}_1^B (\vec{u}^2,0) 
+ \vec{u}^2 {\partial{\cal D}_1^B \over \partial \vec{u}^2} (\vec{u}^2,0) ~; 
\nonumber \\
{\cal D}_\perp^B (\vec{u}^2,0) &\equiv& {\cal D}^B (\vec{u}^2,0) + 
{\cal D}_1^B (\vec{u}^2,0) ~, 
\ea
by measuring appropriate linear superpositions of the correlators (2.7) 
and (2.8) at equal times ($u_4 = 0$), on a $16^3 \times 4$ lattice 
(so that, in our notation, $N_\tau = 4$). 
The critical temperature $T_c$ for such a lattice corresponds to
$\beta_c \simeq 5.69$.
The behaviour of ${\cal D}_\parallel^E$ and ${\cal D}_\perp^E$ is shown 
in Figs. 1 and 2 respectively, on three--dimensional plots versus
$T/T_c$ and the physical distance in the range $0.4$ up to 1 fm.
Due to the logarithmic scale, the errors are comparable with
the size of the symbols and the lines connecting the points are drawn as an 
eye--guide. A clear drop is observed for ${\cal D}_\parallel^E$ and 
${\cal D}_\perp^E$ across the phase transition, as expected.

The analogous behaviour for ${\cal D}_\parallel^B$ and ${\cal D}_\perp^B$ 
is shown in Figs. 3 and 4. No drop is visible across the transition for 
the magnetic correlations.

Finally, we have also measured the mixed electric--magnetic correlator
of Eq. (2.9). When computed at equal times ($u_4 = 0$), this 
correlator turns out to be zero both at zero temperature and at finite 
temperature. This is because the function $D_1^{BE}$ at the right--hand 
side of Eq. (2.9) depends on $u_4^2$, as a consequence of the invariance of 
the theory under time--reversal. We have verified this directly on the 
lattice. Then, we have measured the same correlator (2.9) for $u_4 = 1$
lattice spacing and $|\vec{u}| = 3,4,5,6,7,8$ lattice spacings. 
The results of these measurements are shown in Fig. 5.

Five tables with the values of the various correlators together with the errors
(including an estimate of the systematic uncertainties) are included for the 
convenience of the interested reader.

Our results can be summarized as follows:

(i) In the confined phase ($T < T_c$), until very near to the temperature of 
deconfinement, the correlators, both the electric--electric type (2.7) and
the magnetic--magnetic type (2.8), are nearly equal to the correlators at 
zero temperature \cite{DiGiacomo96}: in other words, $D^E \simeq D^B
\simeq D$ and $D_1^E \simeq D_1^B \simeq D_1$ for $T < T_c$.

(ii) Immediately above $T_c$, the electric--electric correlators (2.7) have 
a clear drop, while the magnetic--magnetic correlators (2.8) stay almost 
unchanged, or show a slight increase.

\bigskip
\noindent {\bf Acknowledgements}
\smallskip

This work was done using the CRAY T3D of the CINECA Inter University 
Computing Centre (Bologna, Italy). We would like to thank the CINECA for 
having put the CRAY T3D at our disposal and for the kind and highly qualified 
technical assistance. 

Stimulating discussions with Yuri Simonov are warmly acknowledged.

\vfill\eject

{\renewcommand{\Large}{\normalsize}
}

\vfill\eject

\pagestyle{empty}

\noindent
\begin{center}
{\bf FIGURE CAPTIONS}
\end{center}
\vskip 0.5 cm
\begin{itemize}
\item [\bf Fig.~1.] The quantity ${\cal D}_\parallel^E / \Lambda_L^4$
[defined by the first Eq. (3.1)] versus $T/T_c$ and the physical 
distance (in fm).
\bigskip
\item [\bf Fig.~2.] The quantity ${\cal D}_\perp^E / \Lambda_L^4$
[defined by the second Eq. (3.1)] versus $T/T_c$ and the physical 
distance (in fm).
\bigskip
\item [\bf Fig.~3.] The quantity ${\cal D}_\parallel^B / \Lambda_L^4$
[defined by the third Eq. (3.1)] versus $T/T_c$ and the physical 
distance (in fm).
\bigskip
\item [\bf Fig.~4.] The quantity ${\cal D}_\perp^B / \Lambda_L^4$
[defined by the fourth Eq. (3.1)] versus $T/T_c$ and the physical 
distance (in fm).
\bigskip
\item [\bf Fig.~5.] The quantity $-(1/2) (|\vec{u}| / \Lambda_L^4)
(\partial {\cal D}_1^{BE} / \partial u_4)$, appearing at the right--hand 
side of Eq. (2.9), for $u_4 = 1$ lattice spacing (and 
$|\vec{u}| = 3,4,5,6,7,8$ lattice spacings), versus $T/T_c$ and the 
physical spatial distance $|\vec{u}|$ (in fm).
\end{itemize}

\vfill\eject

\begin{table}[hbt]
\centering
\small
\setlength{\tabcolsep}{1.5pc}
\caption{The values of the correlator plotted in Fig. 1.}
\vspace{0.3cm}
\label{tab:table1}
\begin{tabular}{rrrr}
\hline
$T/T_c$ & $d_{\rm phys}$ & correlator & error \\
& & & \\
\hline
0.956 & 4.19942108E-01 & 7.39002927E+07 & 2.73704788E+06 \\
0.956 & 5.59922811E-01 & 3.07028346E+07 & 3.07917886E+05 \\
0.956 & 6.99903514E-01 & 1.38699901E+07 & 2.32649070E+05 \\
0.956 & 8.39884217E-01 & 5.95307914E+06 & 1.36852394E+05 \\
0.956 & 9.79864920E-01 & 1.71065492E+06 & 1.36852394E+05 \\
0.978 & 4.10624802E-01 & 7.55998203E+07 & 2.24553922E+06 \\
0.978 & 5.47499737E-01 & 3.18043204E+07 & 3.29345752E+05 \\
0.978 & 6.84374671E-01 & 1.40046796E+07 & 2.76949837E+05 \\
0.978 & 8.21249605E-01 & 5.94319379E+06 & 1.72158007E+05 \\
0.978 & 9.58124539E-01 & 1.33983840E+06 & 1.64672876E+05 \\
1.011 & 3.97031115E-01 & 3.90523140E+07 & 1.88410287E+06 \\
1.011 & 5.29374819E-01 & 1.07736428E+07 & 8.30718082E+05 \\
1.011 & 6.61718524E-01 & 2.72338505E+06 & 3.85384677E+05 \\
1.034 & 3.88217074E-01 & 1.01557232E+07 & 5.80862394E+05 \\
1.034 & 5.17622766E-01 & 1.82690592E+06 & 1.68637469E+05 \\
1.070 & 3.75357913E-01 & 7.16100968E+06 & 3.10882157E+05 \\
1.070 & 5.00477217E-01 & 1.22530450E+06 & 1.06128736E+05 \\
1.131 & 3.54856502E-01 & 7.59600890E+06 & 2.81830719E+05 \\
1.131 & 4.73142002E-01 & 1.42257411E+06 & 9.39435730E+04 \\
\hline
\end{tabular}
\end{table}

\vfill\eject

\begin{table}[hbt]
\centering
\small
\setlength{\tabcolsep}{1.5pc}
\caption{The values of the correlator plotted in Fig. 2.}
\vspace{0.3cm}
\label{tab:table2}
\begin{tabular}{rrrr}
\hline
$T/T_c$ & $d_{\rm phys}$ & correlator & error \\
& & & \\
\hline
0.956 & 4.19942108E-01 & 1.00586510E+08 & 4.78983379E+06 \\
0.956 & 5.59922811E-01 & 3.96598238E+07 & 4.10557182E+05 \\
0.956 & 6.99903514E-01 & 1.89266861E+07 & 1.91593352E+05 \\
0.956 & 8.39884217E-01 & 1.00997067E+07 & 1.23167155E+05 \\
0.956 & 9.79864920E-01 & 5.63147601E+06 & 1.09481915E+05 \\
0.956 & 1.11984562E+00 & 3.18181816E+06 & 1.02639295E+05 \\
0.978 & 4.10624802E-01 & 1.09282908E+08 & 3.74256536E+07 \\
0.978 & 5.47499737E-01 & 4.22161373E+07 & 7.48513072E+05 \\
0.978 & 6.84374671E-01 & 2.01649422E+07 & 2.09583660E+05 \\
0.978 & 8.21249605E-01 & 1.11004489E+07 & 1.42217484E+05 \\
0.978 & 9.58124539E-01 & 6.38481650E+06 & 1.27247222E+05 \\
0.978 & 1.09499947E+00 & 3.75753562E+06 & 1.12276961E+05 \\
1.011 & 3.97031115E-01 & 1.18184634E+08 & 2.56923118E+07 \\
1.011 & 5.29374819E-01 & 3.93948781E+07 & 4.28205197E+06 \\
1.011 & 6.61718524E-01 & 1.56294897E+07 & 3.51128262E+05 \\
1.011 & 7.94062229E-01 & 7.23838065E+06 & 1.74707720E+05 \\
1.011 & 9.26405934E-01 & 3.76820573E+06 & 1.37025663E+05 \\
1.011 & 1.05874964E+00 & 2.23523113E+06 & 1.11333351E+05 \\
1.034 & 3.88217074E-01 & 1.17109354E+08 & 2.81062449E+07 \\
1.034 & 5.17622766E-01 & 3.56012435E+07 & 1.59268721E+07 \\
1.034 & 6.47028457E-01 & 1.26478102E+07 & 2.15481211E+06 \\
1.034 & 7.76434149E-01 & 5.49477087E+06 & 1.40531224E+05 \\
1.034 & 9.05839840E-01 & 2.44992768E+06 & 8.43187346E+04 \\
1.034 & 1.03524553E+00 & 1.22543228E+06 & 5.52756149E+04 \\
1.070 & 3.75357913E-01 & 1.28640892E+08 & 3.21602231E+07 \\
1.070 & 5.00477217E-01 & 4.04146804E+07 & 1.60801116E+07 \\
1.070 & 6.25596522E-01 & 1.45471409E+07 & 1.50081041E+06 \\
1.070 & 7.50715826E-01 & 6.02468180E+06 & 2.25121562E+05 \\
1.070 & 8.75835130E-01 & 2.66822651E+06 & 1.39360967E+05 \\
1.070 & 1.00095443E+00 & 1.32928922E+06 & 1.07200744E+05 \\
1.131 & 3.54856502E-01 & 1.52993819E+08 & 4.02615313E+07 \\
1.131 & 4.73142002E-01 & 4.79112222E+07 & 1.87887146E+07 \\
1.131 & 5.91427503E-01 & 1.65072278E+07 & 1.47625615E+06 \\
1.131 & 7.09713003E-01 & 6.41500398E+06 & 2.68410209E+05 \\
1.131 & 8.27998504E-01 & 2.75120464E+06 & 9.39435730E+04 \\
1.131 & 9.46284005E-01 & 1.31923617E+06 & 7.24707563E+04 \\
\hline
\end{tabular}
\end{table}

\vfill\eject

\begin{table}[hbt]
\centering
\small
\setlength{\tabcolsep}{1.5pc}
\caption{The values of the correlator plotted in Fig. 3.}
\vspace{0.3cm}
\label{tab:table3}
\begin{tabular}{rrrr}
\hline
$T/T_c$ & $d_{\rm phys}$ & correlator & error \\
& & & \\
\hline
0.956 & 4.19942108E-01 & 7.21896378E+07 & 1.36852394E+06 \\
0.956 & 5.59922811E-01 & 3.04428150E+07 & 2.12121211E+05 \\
0.956 & 6.99903514E-01 & 1.34046920E+07 & 1.77908112E+05 \\
0.956 & 8.39884217E-01 & 6.09472137E+06 & 1.02639295E+05 \\
0.956 & 9.79864920E-01 & 2.50371455E+06 & 9.57966758E+04 \\
0.956 & 1.11984562E+00 & 9.85337237E+05 & 8.21114364E+04 \\
0.978 & 4.10624802E-01 & 7.40279428E+07 & 1.34732353E+06 \\
0.978 & 5.47499737E-01 & 3.18342609E+07 & 2.69464706E+05 \\
0.978 & 6.84374671E-01 & 1.45286387E+07 & 1.72158007E+05 \\
0.978 & 8.21249605E-01 & 7.08093366E+06 & 1.04791830E+05 \\
0.978 & 9.58124539E-01 & 3.14375490E+06 & 9.73066993E+04 \\
0.978 & 1.09499947E+00 & 1.31438895E+06 & 8.98215686E+04 \\
1.011 & 3.97031115E-01 & 7.46961146E+07 & 2.82615430E+06 \\
1.011 & 5.29374819E-01 & 3.34256977E+07 & 2.99743638E+05 \\
1.011 & 6.61718524E-01 & 1.61827308E+07 & 1.76420541E+05 \\
1.011 & 7.94062229E-01 & 8.65830908E+06 & 1.06194889E+05 \\
1.011 & 9.26405934E-01 & 4.48245200E+06 & 9.50615537E+04 \\
1.011 & 1.05874964E+00 & 2.17100035E+06 & 8.64974498E+04 \\
1.034 & 3.88217074E-01 & 7.55196062E+07 & 1.87374966E+06 \\
1.034 & 5.17622766E-01 & 3.53763936E+07 & 5.62124898E+05 \\
1.034 & 6.47028457E-01 & 1.79130467E+07 & 2.24849959E+05 \\
1.034 & 7.76434149E-01 & 9.78097322E+06 & 1.03056231E+05 \\
1.034 & 9.05839840E-01 & 5.35236590E+06 & 1.12424980E+05 \\
1.034 & 1.03524553E+00 & 2.69351513E+06 & 1.40531224E+05 \\
1.070 & 3.75357913E-01 & 8.00146351E+07 & 2.35841636E+06 \\
1.070 & 5.00477217E-01 & 3.78633027E+07 & 5.36003718E+05 \\
1.070 & 6.25596522E-01 & 1.93711744E+07 & 2.46561710E+05 \\
1.070 & 7.50715826E-01 & 1.10309565E+07 & 1.34000930E+05 \\
1.070 & 8.75835130E-01 & 6.22836321E+06 & 1.17920818E+05 \\
1.070 & 1.00095443E+00 & 3.24282250E+06 & 1.39360967E+05 \\
1.131 & 3.54856502E-01 & 9.55942958E+07 & 2.68410209E+06 \\
1.131 & 4.73142002E-01 & 4.62336584E+07 & 5.36820417E+05 \\
1.131 & 5.91427503E-01 & 2.41166572E+07 & 2.81830719E+05 \\
1.131 & 7.09713003E-01 & 1.36218181E+07 & 1.61046125E+05 \\
1.131 & 8.27998504E-01 & 7.80805297E+06 & 1.34205104E+05 \\
1.131 & 9.46284005E-01 & 4.16035823E+06 & 2.68410209E+05 \\
\hline
\end{tabular}
\end{table}

\vfill\eject

\begin{table}[hbt]
\centering
\small
\setlength{\tabcolsep}{1.5pc}
\caption{The values of the correlator plotted in Fig. 4.}
\vspace{0.3cm}
\label{tab:table4}
\begin{tabular}{rrrr}
\hline
$T/T_c$ & $d_{\rm phys}$ & correlator & error \\
& & & \\
\hline
0.956 & 4.19942108E-01 & 1.01236558E+08 & 5.47409576E+06 \\
0.956 & 5.59922811E-01 & 3.92971649E+07 & 2.80547408E+05 \\
0.956 & 6.99903514E-01 & 1.80439881E+07 & 2.32649070E+05 \\
0.956 & 8.39884217E-01 & 9.08699896E+06 & 1.43695014E+05 \\
0.956 & 9.79864920E-01 & 4.83499508E+06 & 1.23167155E+05 \\
0.956 & 1.11984562E+00 & 2.54545453E+06 & 1.09481915E+05 \\
0.978 & 4.10624802E-01 & 1.12247020E+08 & 3.44316013E+07 \\
0.978 & 5.47499737E-01 & 4.26502748E+07 & 5.98810457E+05 \\
0.978 & 6.84374671E-01 & 1.96709235E+07 & 2.61979575E+05 \\
0.978 & 8.21249605E-01 & 1.01648075E+07 & 1.52696667E+05 \\
0.978 & 9.58124539E-01 & 5.45740881E+06 & 1.19762091E+05 \\
0.978 & 1.09499947E+00 & 2.95138704E+06 & 1.12276961E+05 \\
1.011 & 3.97031115E-01 & 1.31030790E+08 & 3.59692365E+07 \\
1.011 & 5.29374819E-01 & 4.90980079E+07 & 1.88410287E+06 \\
1.011 & 6.61718524E-01 & 2.31659012E+07 & 2.99743638E+05 \\
1.011 & 7.94062229E-01 & 1.19259429E+07 & 1.71282079E+05 \\
1.011 & 9.26405934E-01 & 6.54922721E+06 & 1.33600021E+05 \\
1.011 & 1.05874964E+00 & 3.69541085E+06 & 1.11333351E+05 \\
1.034 & 3.88217074E-01 & 1.44559786E+08 & 3.74749932E+07 \\
1.034 & 5.17622766E-01 & 5.48165463E+07 & 1.31162476E+06 \\
1.034 & 6.47028457E-01 & 2.63477308E+07 & 4.68437415E+05 \\
1.034 & 7.76434149E-01 & 1.38001662E+07 & 1.87374966E+05 \\
1.034 & 9.05839840E-01 & 7.85663232E+06 & 1.31162476E+05 \\
1.034 & 1.03524553E+00 & 4.70779602E+06 & 1.02119356E+05 \\
1.070 & 3.75357913E-01 & 1.64445941E+08 & 3.64482529E+07 \\
1.070 & 5.00477217E-01 & 6.17476284E+07 & 2.25121562E+06 \\
1.070 & 6.25596522E-01 & 2.97696465E+07 & 3.85922677E+05 \\
1.070 & 7.50715826E-01 & 1.58121097E+07 & 1.82241264E+05 \\
1.070 & 8.75835130E-01 & 9.24392013E+06 & 1.39360967E+05 \\
1.070 & 1.00095443E+00 & 5.68163942E+06 & 1.11488773E+05 \\
1.131 & 3.54856502E-01 & 2.04354112E+08 & 4.16035823E+07 \\
1.131 & 4.73142002E-01 & 7.81744732E+07 & 7.38128073E+06 \\
1.131 & 5.91427503E-01 & 3.80203060E+07 & 1.20784594E+06 \\
1.131 & 7.09713003E-01 & 2.00234016E+07 & 2.08017912E+05 \\
1.131 & 8.27998504E-01 & 1.16269934E+07 & 1.67756380E+05 \\
1.131 & 9.46284005E-01 & 7.08810968E+06 & 1.28836900E+05 \\
\hline
\end{tabular}
\end{table}

\vfill\eject

\begin{table}[hbt]
\centering
\small
\setlength{\tabcolsep}{1.5pc}
\caption{The values of the correlator plotted in Fig. 5.}
\vspace{0.3cm}
\label{tab:table5}
\begin{tabular}{rrrr}
\hline
$T/T_c$ & $d_{\rm phys}$ & correlator & error \\
& & & \\
\hline
0.956 & 4.19942108E-01 & 8.24169594E+06 & 4.78983379E+05 \\
0.956 & 5.59922811E-01 & 3.94446234E+06 & 1.78678249E+05 \\
0.956 & 6.99903514E-01 & 1.87118963E+06 & 1.32152198E+05 \\
0.956 & 8.39884217E-01 & 8.86735087E+05 & 1.08077810E+05 \\
0.956 & 9.79864920E-01 & 4.00217984E+05 & 1.02933186E+05 \\
0.956 & 1.11984562E+00 & 1.38701270E+05 & 9.77759035E+04 \\
0.978 & 4.10624802E-01 & 8.57417981E+06 & 5.98810457E+05 \\
0.978 & 5.47499737E-01 & 4.10971102E+06 & 2.17131666E+05 \\
0.978 & 6.84374671E-01 & 1.96096577E+06 & 1.41702507E+05 \\
0.978 & 8.21249605E-01 & 9.86734842E+05 & 1.07524651E+05 \\
0.978 & 9.58124539E-01 & 4.82581348E+05 & 1.02201226E+05 \\
0.978 & 1.09499947E+00 & 2.38981511E+05 & 9.60873715E+04 \\
1.011 & 3.97031115E-01 & 4.46635149E+06 & 5.13846236E+05 \\
1.011 & 5.29374819E-01 & 2.03499810E+06 & 1.92022626E+05 \\
1.011 & 6.61718524E-01 & 8.89660527E+05 & 1.18007786E+05 \\
1.011 & 7.94062229E-01 & 4.74053127E+05 & 9.08201812E+04 \\
1.011 & 9.26405934E-01 & 2.64646227E+05 & 8.28624159E+04 \\
1.011 & 1.05874964E+00 & 1.57385964E+05 & 7.80386843E+04 \\
\hline
\end{tabular}
\end{table}

\vfill\eject

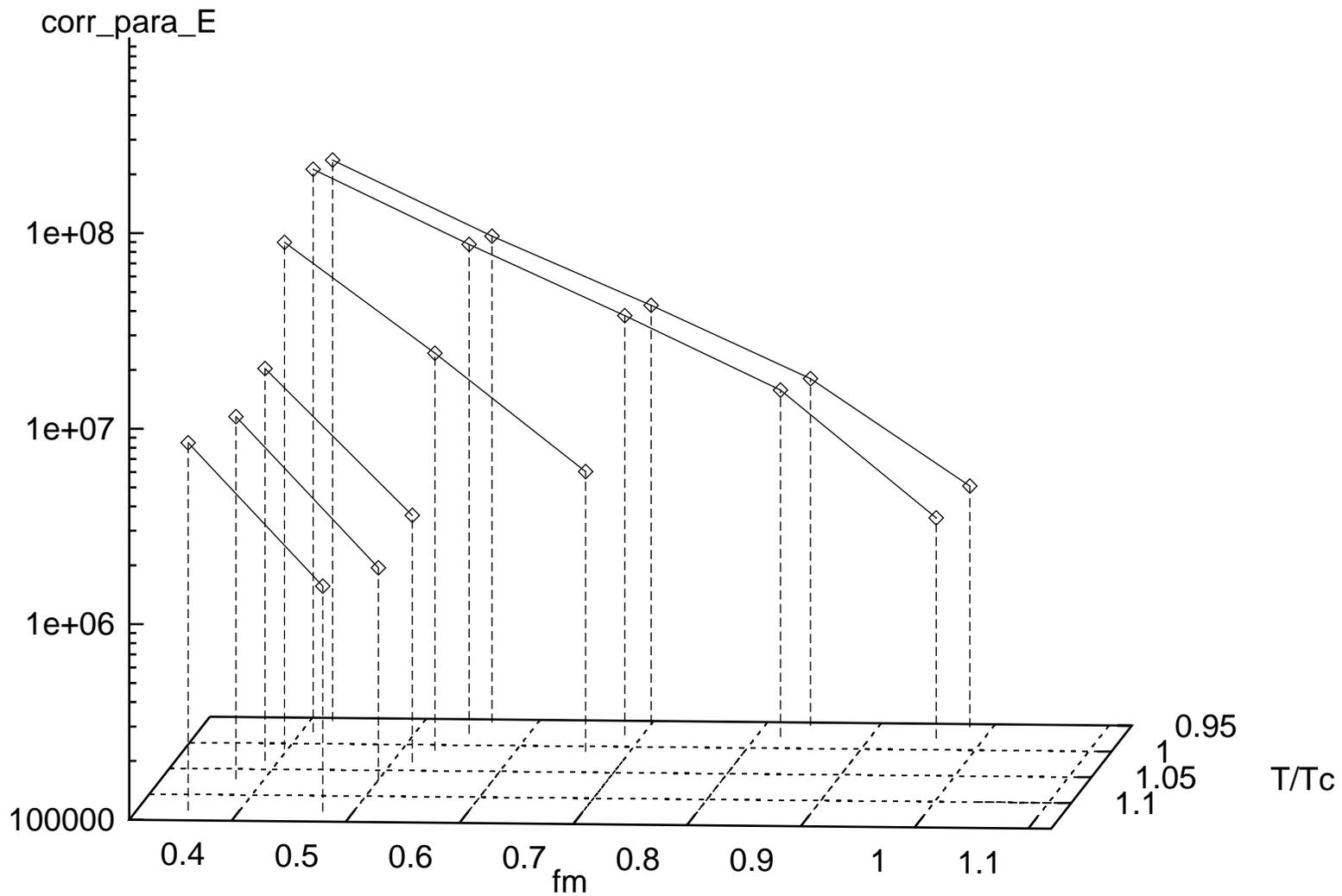
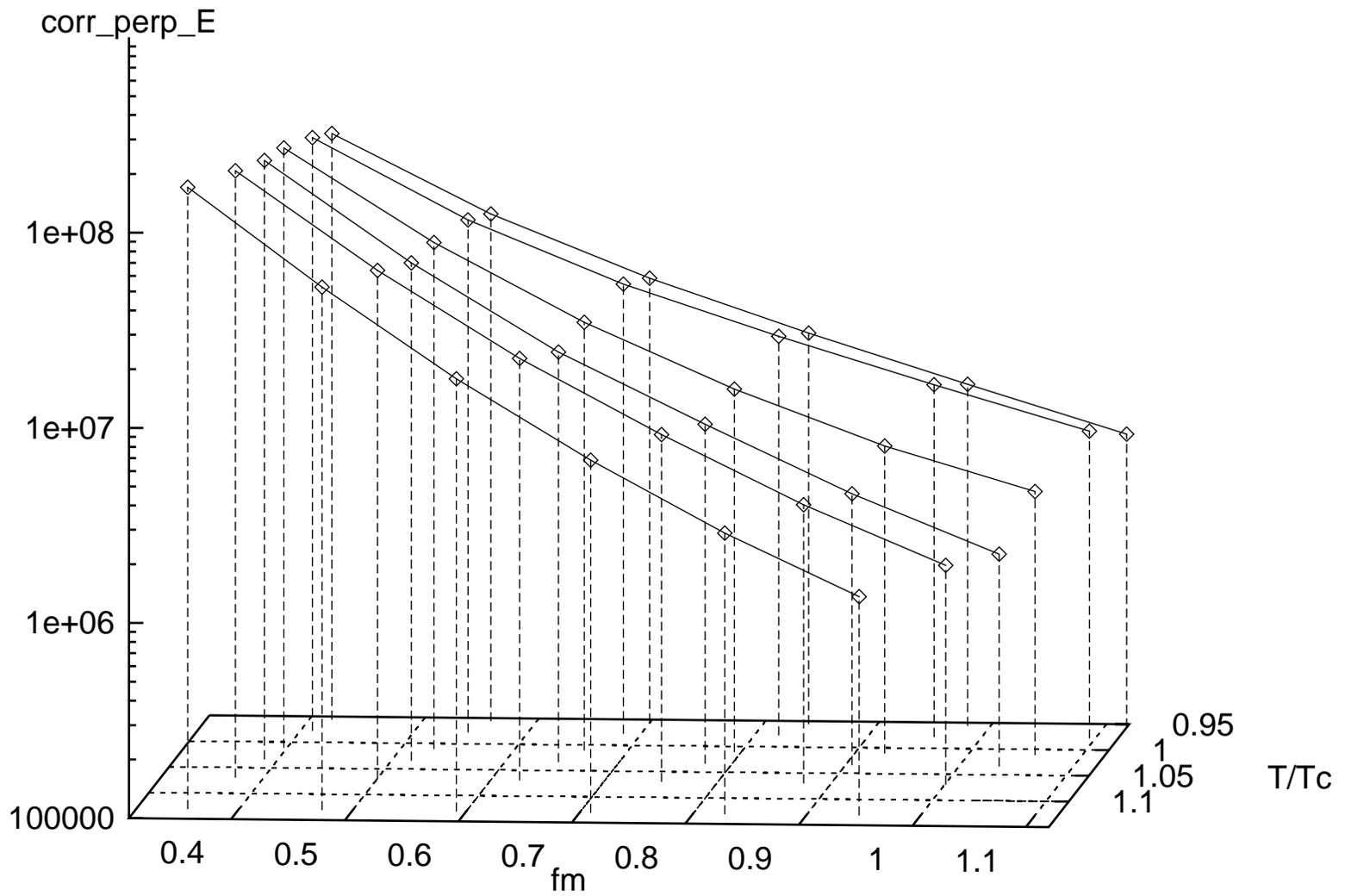
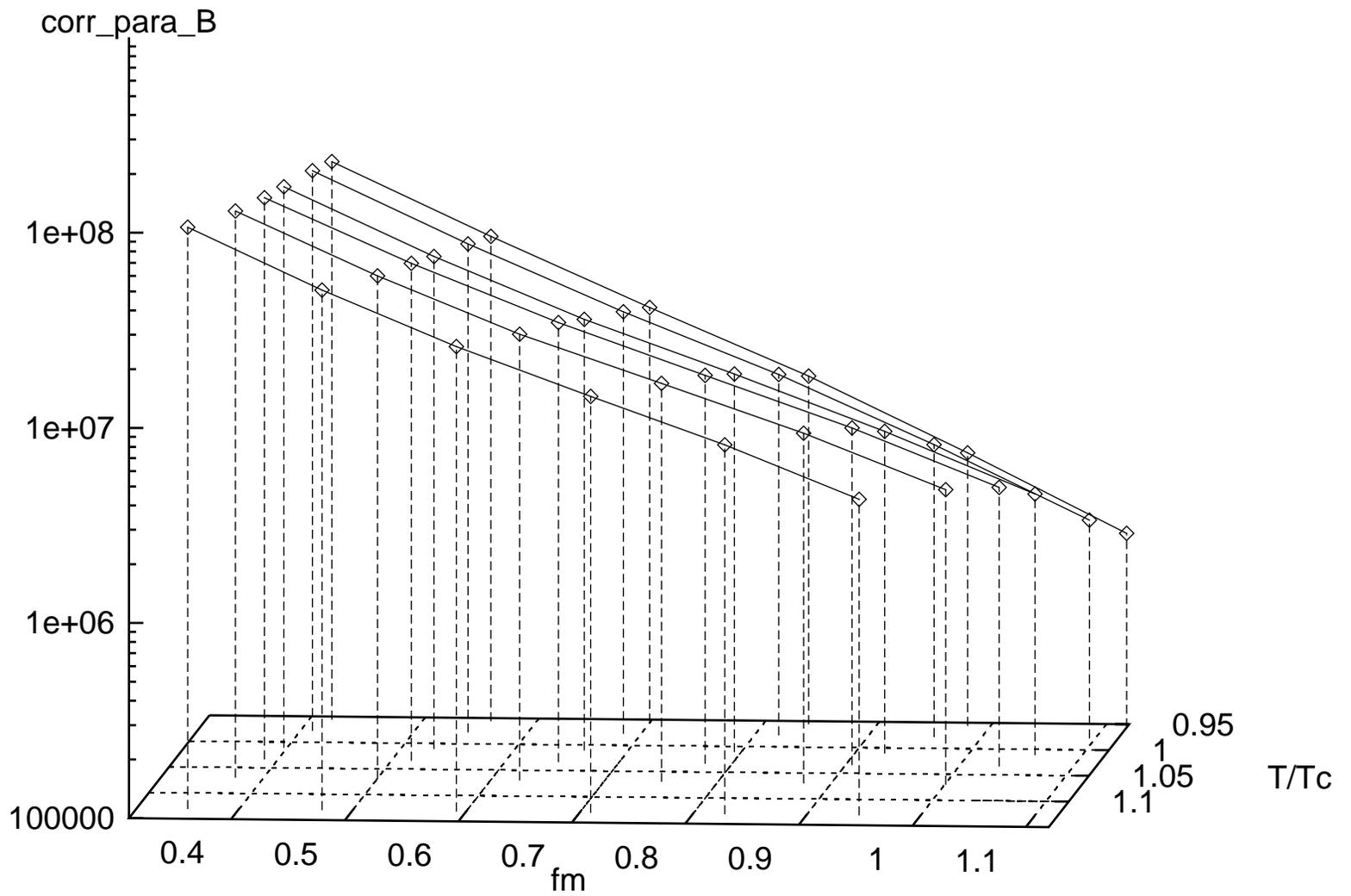
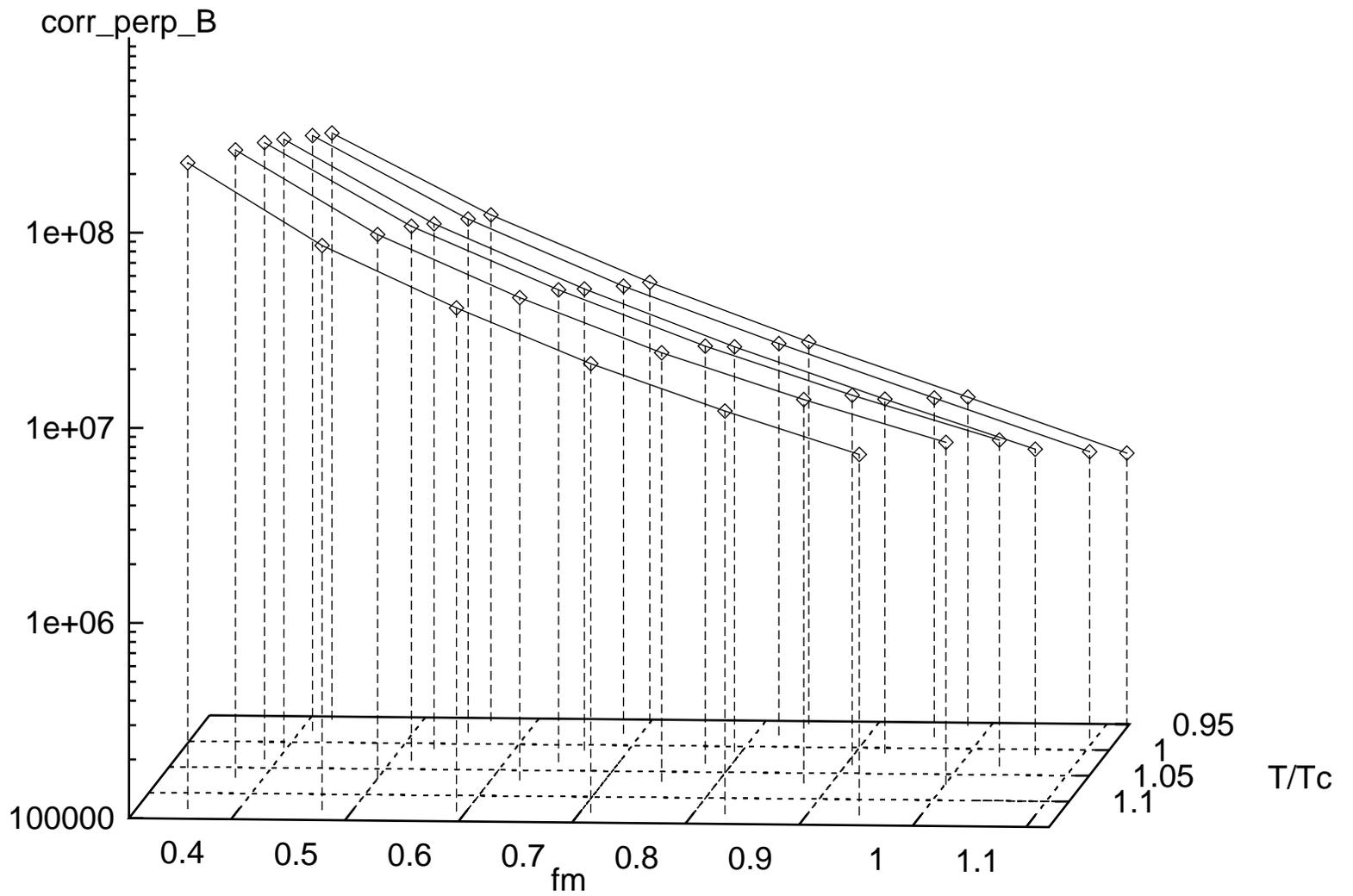
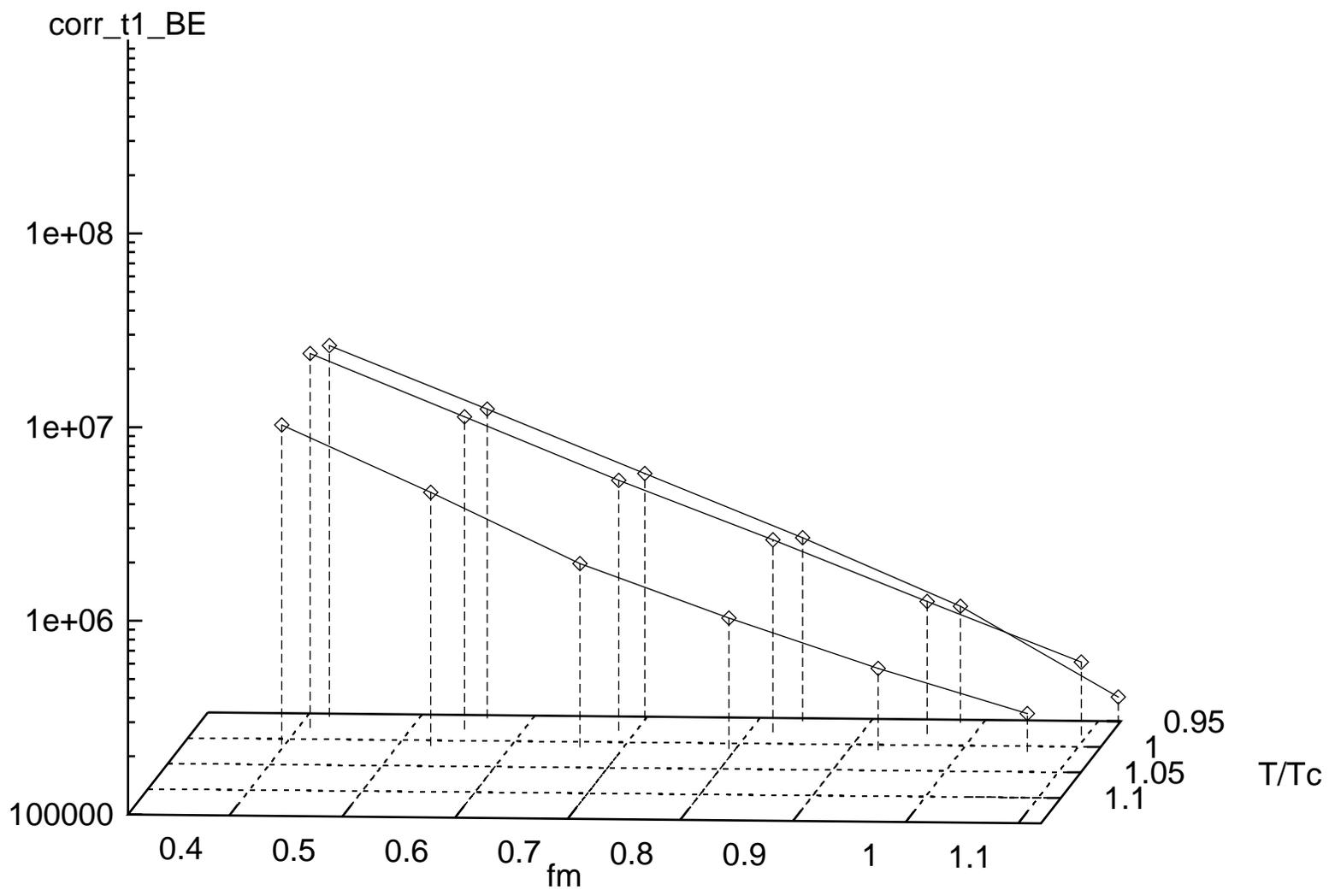
\end{document}